&latex209
\documentclass[11pt]{article}
\usepackage{moriond-photos,epsfig}
\def\be{\begin{equation}}
\def\ee{\end{equation}}
\def\bea{\begin{eqnarray}}
\def\eea{\end{eqnarray}}

\begin{document}
\vspace*{4cm}
\title{HIDDEN AND OPEN BEAUTY PRODUCTION IN 920 GeV PROTON--NUCLEUS COLLISIONS}

\author{Martin zur Nedden for the HERA-$B$ collaboration}

\address{Institut f\"ur Physik, Humboldt Universit\"at zu Berlin,\\
Newtonstra{\ss}e 15, D--12489 Berlin, Germany}

\maketitle\abstracts{
  The $b\bar{b}$ and $\Upsilon$ production cross section has been measured in collisions of $920~\mbox{GeV}$ protons on different 
  nuclear targets using the HERA-$B$ detector. On the invariant mass spectra of $e^{+}e^{-}$ or $\mu^{+}\mu^{-}$
  events originating from common vertex, the $\Upsilon$ search was performed. The  identification of $b\bar{b}$ events was done 
  using the same sample via inclusive bottom quark decays into $J/\psi$ by  exploiting the longitudinal separation of the 
  $J/\psi \rightarrow l^{+}l^{-}$ decay vertices from the primary proton nucleus interaction. For both hidden and open beauty
  decays, the $\mu^{+}\mu^{-}$ and the $e^{+}e^{-}$ decay channels have been reconstructed. The most recent measurement, 
  using data collected in 2002/2003, yields a preliminary cross section in the combined analysis of $\sigma(b\bar{b}) =  
  (12.3^{+3.5}_{-3.2})~\mbox{nb/nucleon}$ and for the hidden beauty production the preliminary measurement of 
  $BR(\Upsilon \rightarrow l^{+}l^{-}) \times d\sigma(\Upsilon \rightarrow l^{+}l^{-})/dy|_{y=0} = (3.4 \pm 0.8)~\mbox{pb/nucleon}$.}

\vspace{-0.2cm}

\section{Introduction}

The HERA-$B$ \cite{herab1} experiment was designed to identify $B$--meson decays in a dense
hadronic environment with a large geometrical coverage. The $B$ mesons are produced by interaction of the protons in the halo of the
$920~\mbox{GeV}$ HERA proton beam with different target wires, which can be used simultaneously. Since there are
various target materials available, the $A$--dependence of heavy quark production can be measured, too. The events from different
wires can be separated easily by applying accurate spatial separation cuts.

In the data taking period of 2002/2003 starting in October 2002 HERA-$B$ was routinely running and collected $164 \cdot 10^{6}$
events triggered by a dilepton $J/\psi$--trigger. HERA-$B$ is able to reconstruct the $J/\psi$ and the $\Upsilon$ either in the 
$\mu^{+}\mu^{-}$ or in the $e^{+}e^{-}$ decay channel. The availability of both channels is important to increase the statistics 
and to cross check the results.

\subsection{Dilepton Event Selection}

The number of prompt $J/\psi$ decays $n_{P}$, which are produced directly at the target wire, 
is used as a normalization factor of the $b\bar{b}$ and $\Upsilon$ cross section 
measurement. The selection of the $J/\psi/\Upsilon \rightarrow l^{+}l^{-}$ differs between the $\mu$-- and $e$--channel due to the different 
shapes and yields of the background. To select the $\mu^{+}\mu^{-}$ decays, a dimuon vertex is required in addition to muon identification 
cuts in the Muon and RICH systems. Aside from the dilepton vertex additional identification cuts based on the electromagnetic calorimeter (ECAL) 
are applied for the electrons. 

During the data taking period of 2002/2003 about 160000 $J/\psi \rightarrow \mu^{+}\mu^{-}$ and approximately 140000 $J/\psi \rightarrow
e^{+}e^{-}$ events could be reconstructed. They are the basic sample of both, the open and hidden beauty measurements described below.

\section{Hidden Beauty Measurements}

\begin{figure}
  \centering
  \includegraphics[height=.16\textheight]{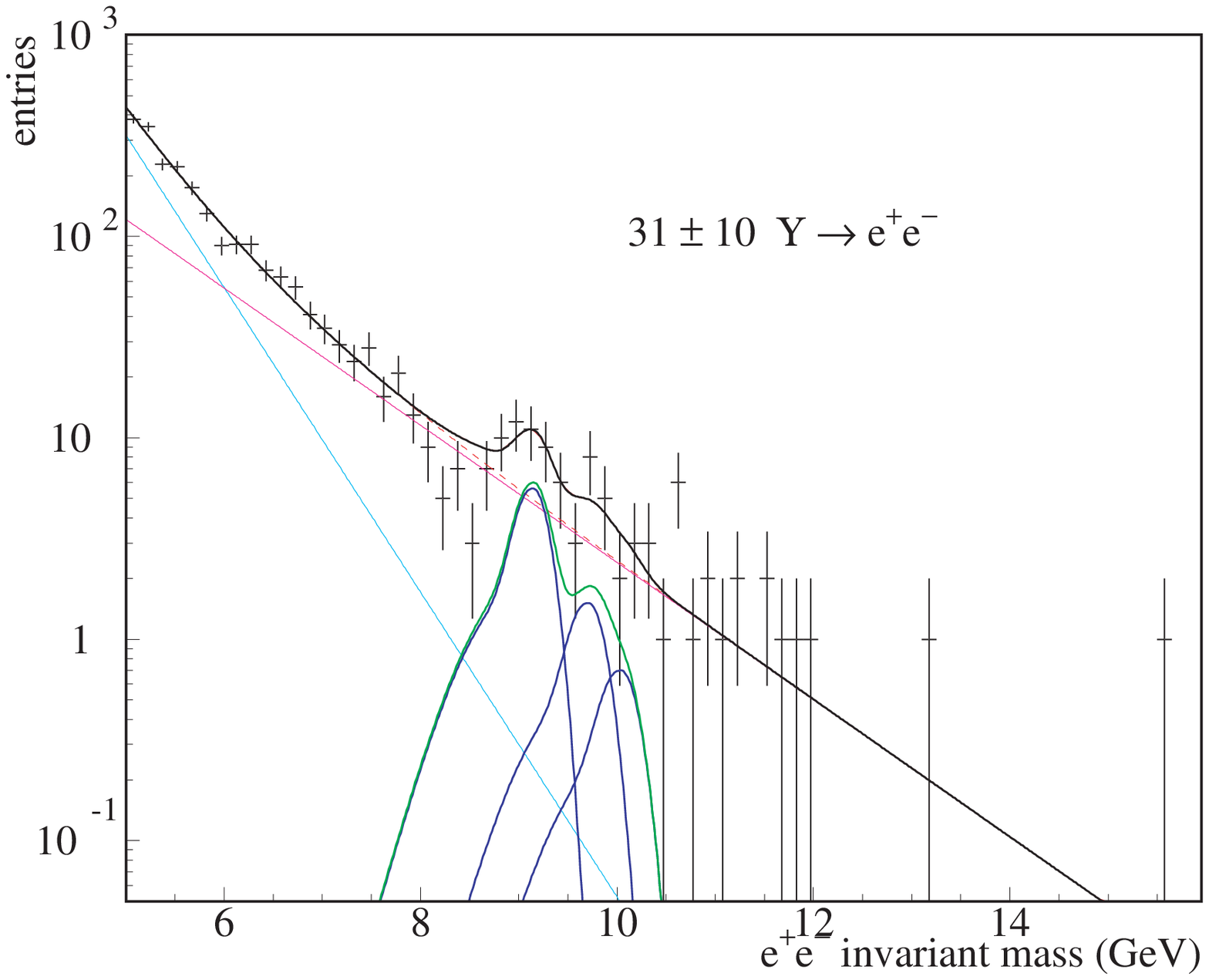}\hspace{2.0cm}
  \includegraphics[height=.16\textheight]{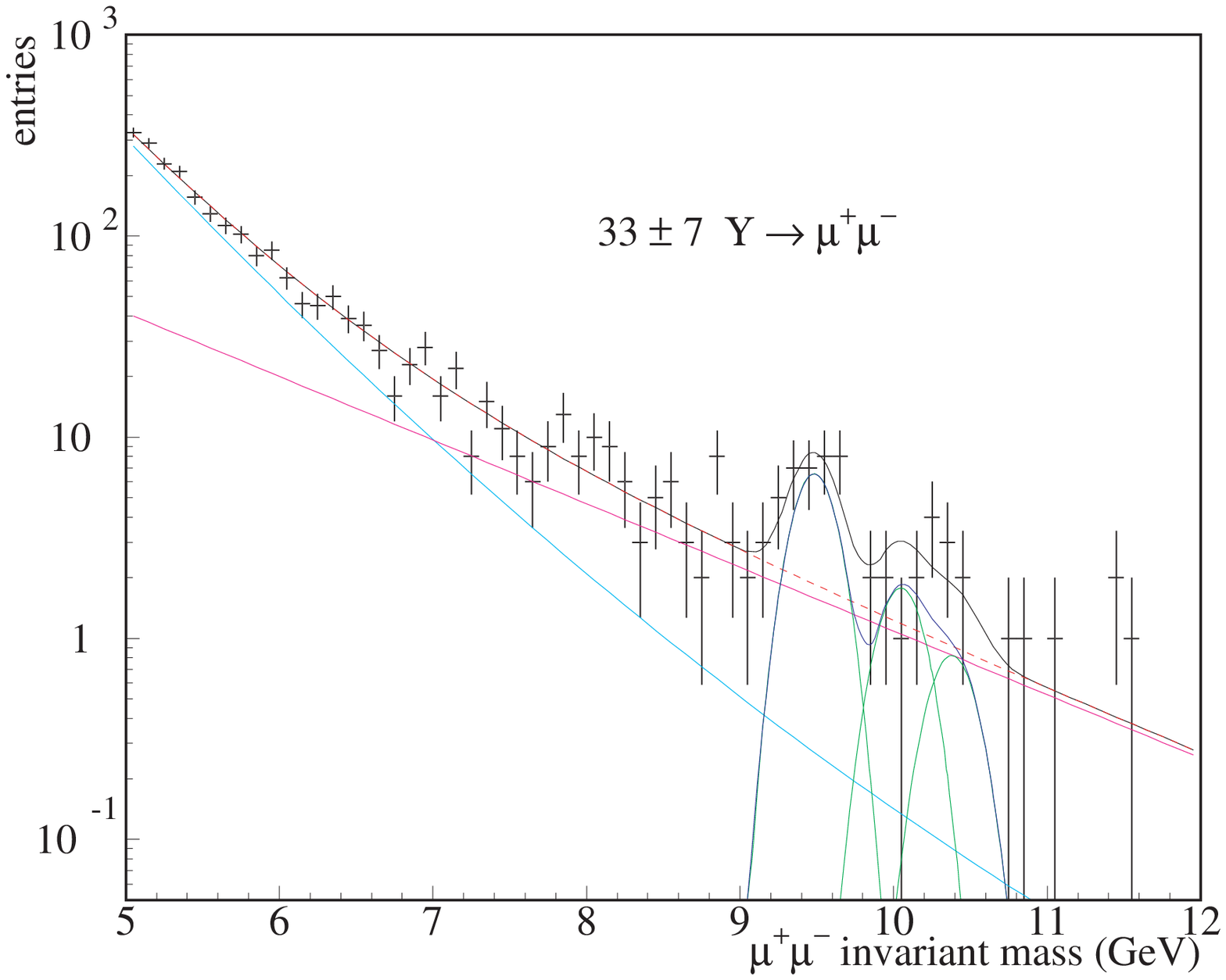}
  \vspace{-0.5cm}
  \caption{The $\Upsilon$ signal in the invariant mass of the dilepton events in the electron and muon channel.}
  \label{fig:upsdec}
\end{figure}

In order to minimize the sensitivity to systematic effects from luminosity and Monte Carlo (MC) efficiency determination, the cross section
analysis is done using a relative measurement with respect to the number of prompt $J/\psi$ events ($n_{P}$). The $\Upsilon$ signal is
clearly visible in both data samples (Fig.~\ref{fig:upsdec}) with a mass resolution between $140 - 160~\mbox{MeV}$. The background
shapes are described by a combination of combinatorial background and Drell-Yan events, whereas the $\Upsilon$ signal
shapes are taken from MC. The relative production fraction of $\Upsilon(1S)/\Upsilon(2S)/\Upsilon(3S)$ is fixed at the
values measured by E605 \cite{e605}. The obtained
results are stable with respect to changes in the selection and background description.
The HERA-$B$ preliminary cross section measurement of $BR(\Upsilon \rightarrow l^{+}l^{-}) \times \frac{d\sigma_{\Upsilon}}{dy}\Big|_{y=0}(\sqrt{s}) 
= (3.4 \pm 0.18_{stat})~\mbox{pb/nucleon}$ is in perfect agreement to other experimental data (Fig.~\ref{fig:final}). 
The cross sections measured individually for the $e^{+}e^{-}$ and $\mu^{+}\mu{-}$ data sample are compatible, ($BR \times 
d\sigma_{\Upsilon}^{e^{+}e^{-}}/dy|_{y=0} = (2.9 \pm 1.2_{stat})~\mbox{pb/nucleon}$, 
$BR \times d\sigma_{\Upsilon}^{\mu^{+}\mu^{-}}/dy|_{y=0} = (3.9 \pm 1.1_{stat})~\mbox{pb/nucleon}$).
Amodified Craigie parameterization is used to account for possible nuclear suppression:
\begin{equation}\label{eq:cragie}
  BR(\Upsilon \rightarrow l^{+}l^{-}) \times \frac{d\sigma_{\Upsilon}}{dy}\Big|_{y=0}(\sqrt{d}) =
  \sigma_{0} \cdot e^{-\frac{m_{0}}{\sqrt{s}}} \cdot A^{\alpha - 1}
\end{equation}
where $\sigma_{0} = 144 \pm 28~\mbox{pb/nucleon}$ and $m_{0} = 161 \pm 9~\mbox{GeV}$. The HERA-$B$ data are well described by
the fit (Fig.~\ref{fig:final}), where the high value at $\sqrt{s} = 38~\mbox{GeV}$ was excluded. All the available measurements done
at $\sqrt{s} < 65~\mbox{GeV}$ are compatible to no nuclear suppression with $\alpha = 0.99 \pm 0.05$. 

\section{Open Beauty Production}

\subsection{Cross Section measurement}

As for the $\Upsilon$ cross section, the $\sigma(b\bar{b})$ cross section is determined in a relative measurement to
$n_{P}$ using the known prompt $J/\psi$ cross section $\sigma_{J/\psi}^{A}$ \cite{bb2000}. Within the
HERA-$B$ acceptance, the first fixed--target experiment covering the negative $x_{F}$ region ($x_{F} \in [-0.35,0.15]$,
$x_{F} = \frac{p^{cms}_{L}}{(p^{cms}_{L})_{max}}$), the cross section can be written as as:
\begin{equation} \label{eq:bcros}
  \Delta\sigma(b\bar{b}) = \Delta\sigma_{r} \cdot \frac{n_{B}}{n_{P}} \cdot \frac{1}{\epsilon_{R} \cdot
    \epsilon^{\Delta z}_{B} \cdot BR(b\bar{b} \rightarrow J/\psi X)}
\end{equation}
where $n_{B}/n_{P}$ are the observed $b$ and prompt $J/\psi$ events, $\epsilon_{R}$ the relative efficiency
($\epsilon_{R} = \epsilon^{J/\psi}_{B}/\epsilon^{tot}_{P}$, $\epsilon^{tot}_{B} = \epsilon^{J/\psi}_{B} \cdot \epsilon^{\Delta z}_{B}$)
of $B$ and prompt $J/\psi$ events (trigger + reconstruction + selection), $BR(b\bar{b} \rightarrow J/\psi) = (2.32 \pm 0.20)\,\%$
is the  branching ratio measured at LEP \cite{brlep} and $\Delta\sigma_{r}$ and $\Delta\sigma(b\bar{b})$ ar the reference (prompt $J/\psi$) and
measured ($b\bar{b}$) cross sections inside the HERA-$B$ acceptance.

The reference prompt $J/\psi$ production cross section per nucleon $\sigma(pN \rightarrow J/\psi X)$ was previously measured by two
fixed target experiments (E789 and E771 \cite{e789dir,e771dir}). After correcting for the most recent
measurement of the nuclear dependence $A^{\alpha}$ using $\alpha = 0.955 \pm 0.005$ (E866 \cite{e866}) and scaling to HERA-$B$
energies \cite{e771scal} a reference cross section of $\sigma(pN \rightarrow J/\psi) = (357 \pm 8 \pm 27)~\mbox{nb}/\mbox{nucleon}$ is
obtained. Within the acceptance of HERA-$B$ only a fraction of $f_{P} = (77 \pm 1)\,\%$ \cite{e789dir} of the prompt $J/\psi$ and
$f_{B} = (90.6 \pm 0.5)\,\%$ of the $b\bar{b}$ events can be measured. Therefore, the reference prompt $J/\psi$ cross section to be used
reads:
\begin{equation}
  \Delta \sigma_{r} = f_{P} \, \cdot \, \sigma(pN \rightarrow J/\psi) \, \cdot \, A^{\alpha}
  \label{eq:cross}
\end{equation}
Since no nuclear suppression has been observed in $D$--Meson production \cite{dsup} and a similar behavior is expected in the $b$
channel, a nuclear dependence of $\alpha = 1.0$ is assumed for the $b\bar{b}$ production cross section \cite{bsup} in this analysis,
i.e. $\sigma(b\bar{b})^{A} = \sigma(pN \rightarrow b\bar{b}) \cdot A$.

\subsection{Detached Analysis}

To identify $b$--hadrons from the decay chain $pA \rightarrow b\bar{b}X~\mbox{with}~b\bar{b} \rightarrow J/\psi X^{\prime} 
\rightarrow (e^{+}e^{-}/\mu^{+}\mu^{-})X^{\prime}$ on top of the prompt $J/\psi$ selection the decay length $\Delta z$, defined as the distance along the
beam axis between the $J/\psi$ decay vertex and the closest target wire, is used. Additionally, cuts on the minimum impact parameter 
of both leptons are applied. The expected decay length of the $B$--meson of $\approx 7~\mbox{mm}$ is large enough compared to
the $\Delta z$ resolution of $\sigma_{\Delta z} \approx 0.5~\mbox{mm}$. 
\begin{figure}
  \centering
  \includegraphics[height=.2\textheight]{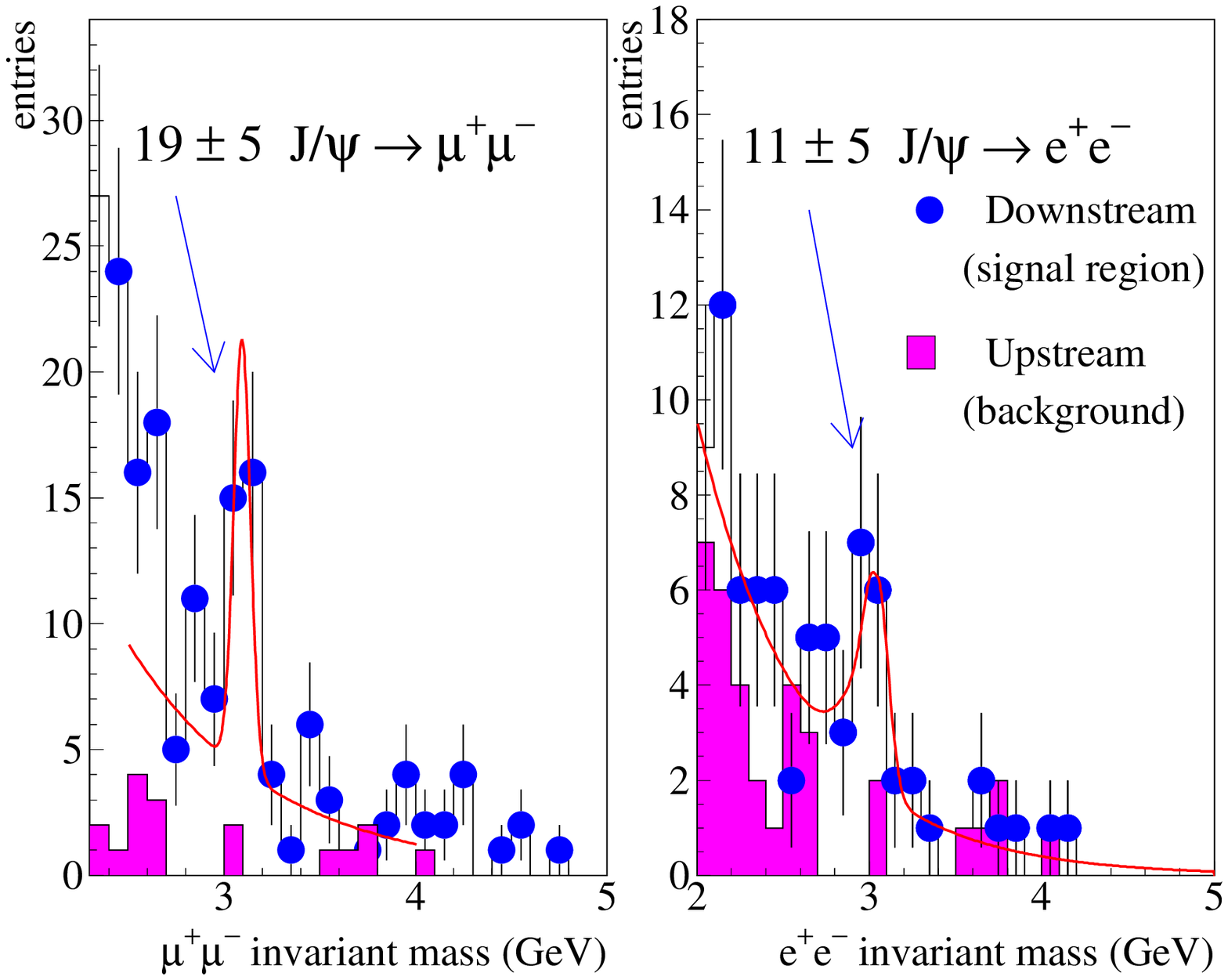} \hspace{2.0cm}
  \includegraphics[height=.18\textheight]{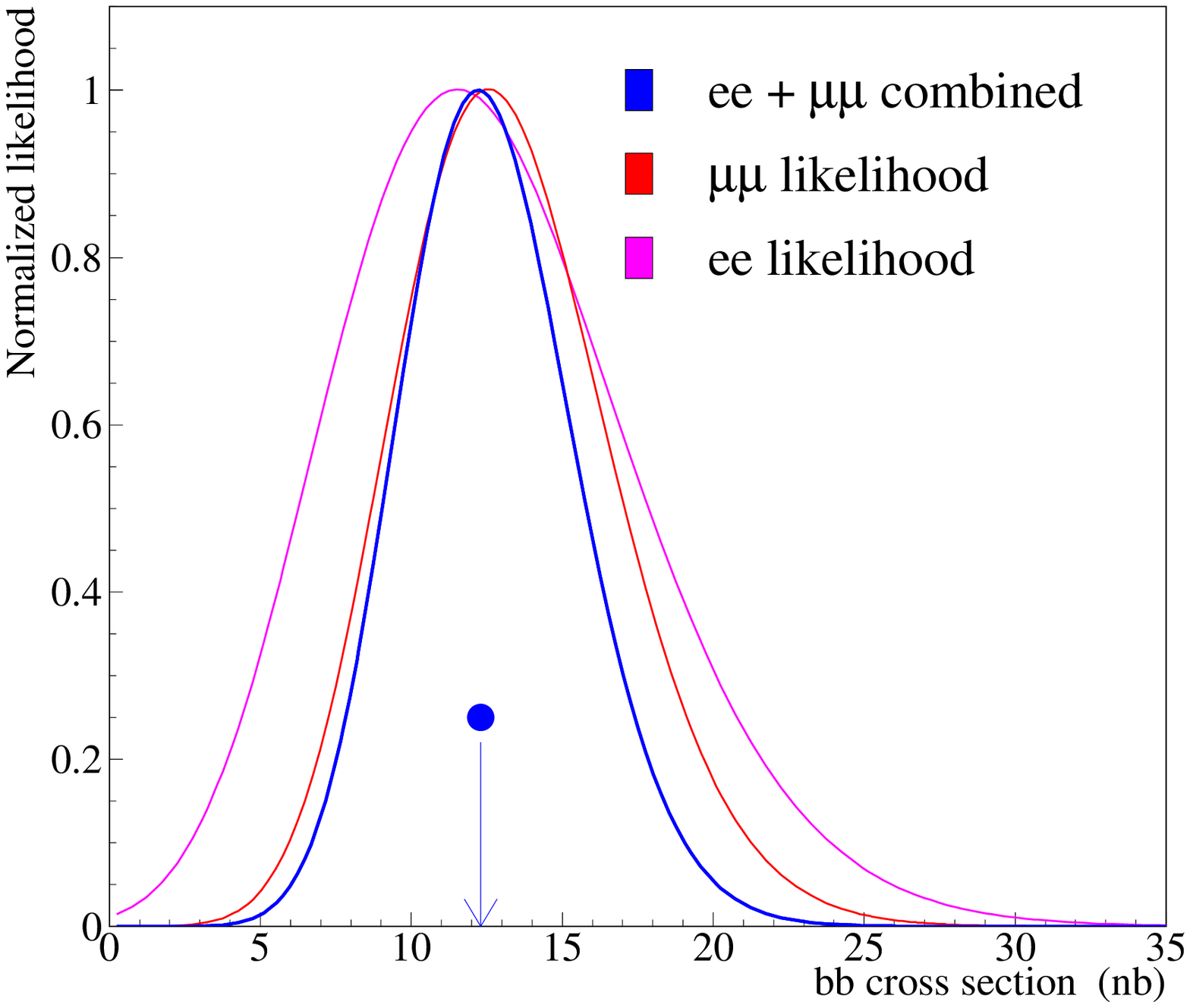}
  \vspace{-0.5cm}
  \caption{The detached $J/\psi$ signal and the maximum likelyhood fit of the $b\bar{b}$ cross section.}
  \label{fig:detjpsi}
\end{figure}
The search for detached $J/\psi$ vertices was performed on $30\,\%$ of the full statistics of the data taken in 2002/2003.
In this amount of data a number of $n_{B}^{\mu} = 19 \pm 5$ detached $J/\psi$'s could be extracted in the $\mu$--channel and 
$n_{B}^{e} = 11 \pm 5$ (Fig.~\ref{fig:detjpsi}) within the $e$--channel were found. The corresponding numbers of the prompt signals are
measured to $n_{P}^{\mu} = 40280 \pm 240$ in the muon channel and $n_{P}^{e} = 35560 \pm 570$ for the electrons. Since
this sample was taken using the carbon wire only, a value of $\Delta\sigma^{C}_{r} = (245 \pm 6 \pm 19) \mbox{nb/nucleon}$ was calculated 
for the corresponding reference prompt $J/\psi$ cross section (Eq.~\ref{eq:cross}). The resulting measurements (Eq.~\ref{eq:bcros}) 
for both decay channels are extracted from a maximum likelihood fit. The likelihood function is shown in Fig.~\ref{fig:detjpsi}
as a function of $\sigma(b\bar{b})$. Both measurements of 
$\sigma(b\bar{b})^{e^{+}e^{-}} = 11.5^{+5.3}_{-4.5}~\mbox{nb/nucleon}$ and 
$\sigma(b\bar{b})^{\mu^{+}\mu^{-}} = 12.7^{+4.1}_{-3.6}~\mbox{nb/nucleon}$ are in good agreement leading to a combined result of
\begin{equation}
  \sigma(b\bar{b}) = 12.3^{+3.5}_{-3.2}~(\mbox{stat.})~\mbox{nb/nucleon}
\end{equation}

\begin{figure}
  \centering
  \includegraphics[height=.18\textheight]{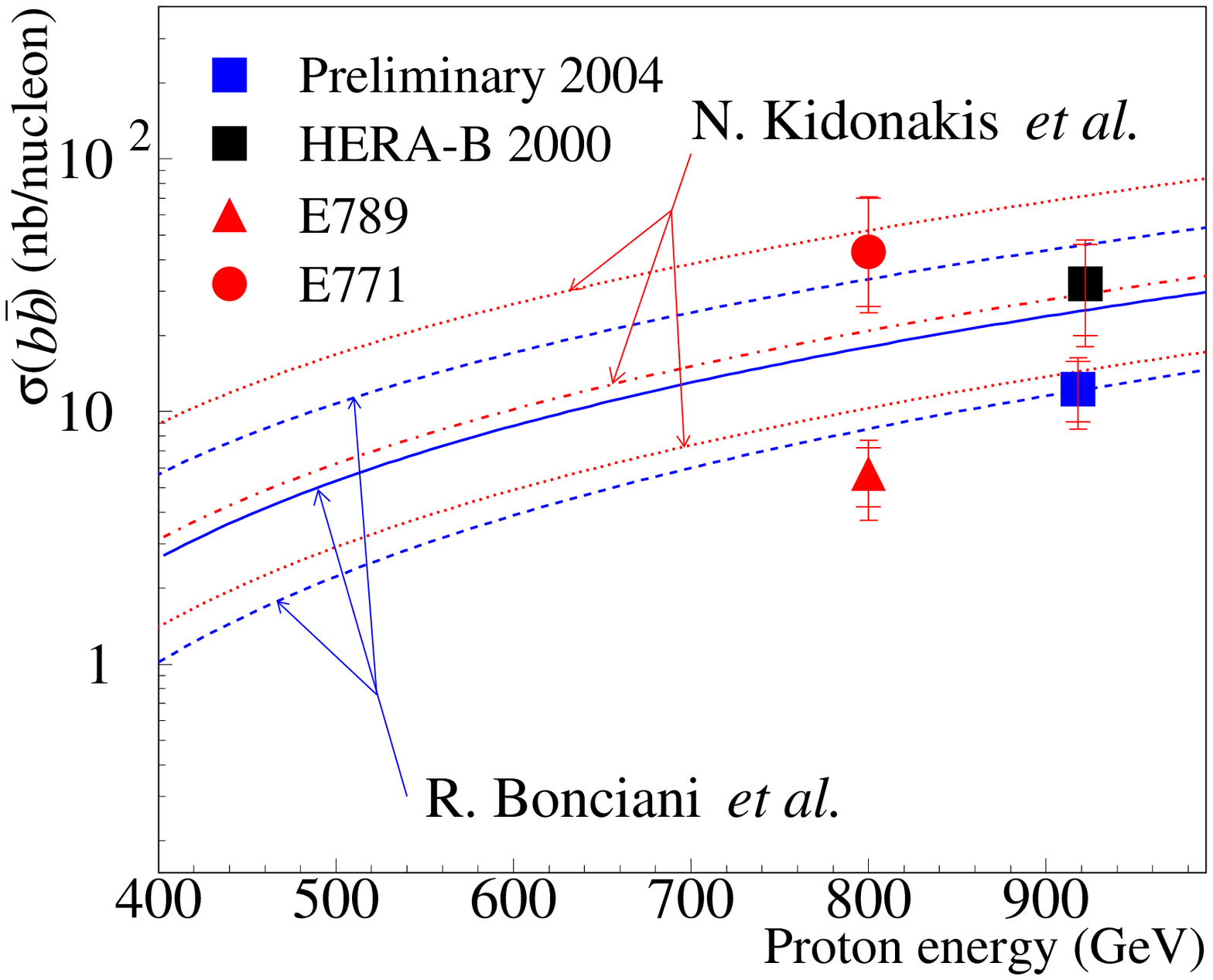}\hspace{2.0cm}
  \includegraphics[height=.18\textheight]{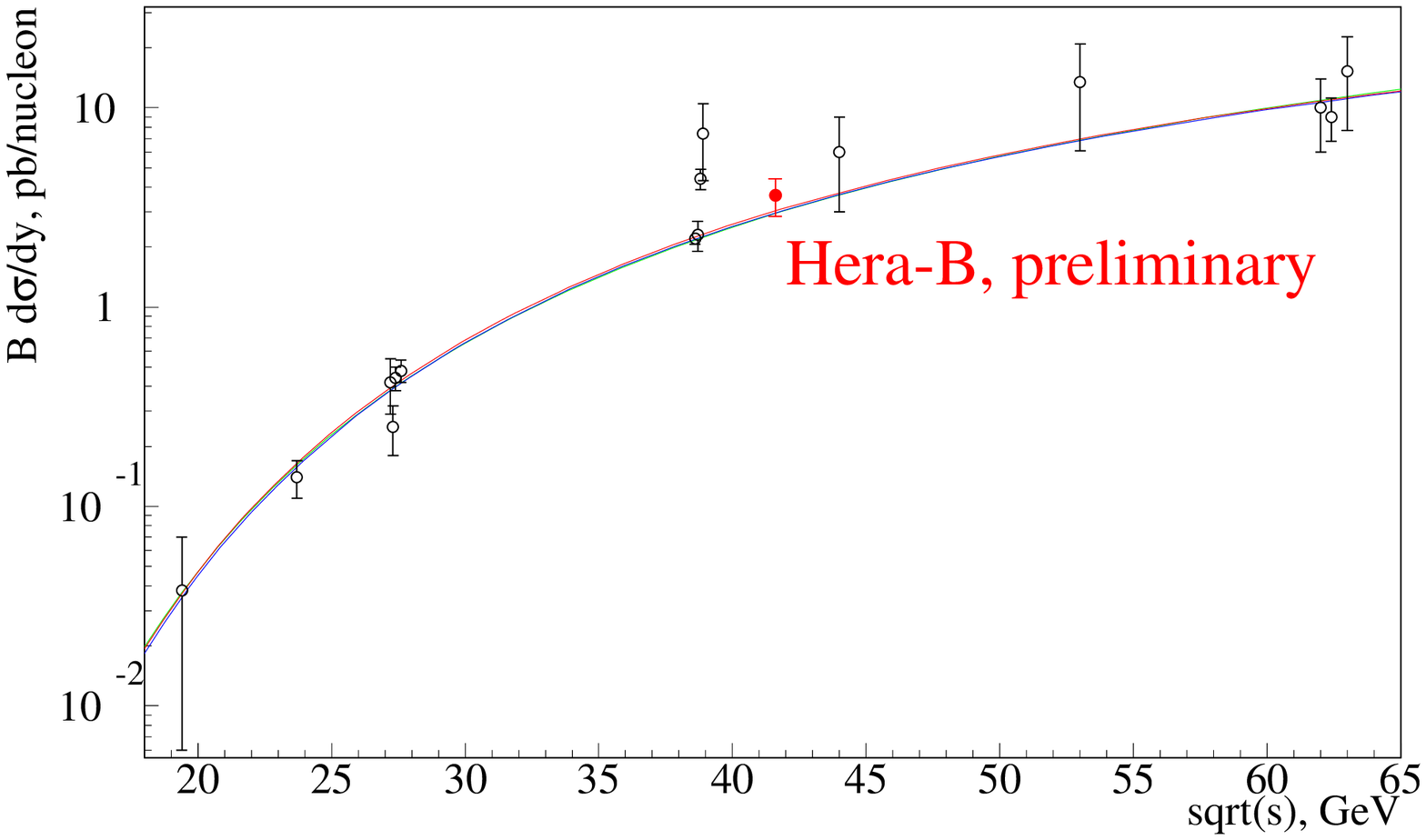}
  \vspace{-0.5cm}
  \caption{Comparsions of $\sigma(b\bar{b})$ measurements to theoretical perdictions and of $BR \times d\sigma_{\Upsilon}/dy$ to the Craigie parametrization.}
  \label{fig:final}
\end{figure}

\vspace{-0.7cm}
       
\section{Conclusions}

Events from open and hidden beauty production have been observed by exploiting the two lepton trigger capabilities of
the HERA-$B$ detector. A measurement of the $\Upsilon$ production cross section is currently going on.
In spite of the limited statistics it will clarify the situation around $\sqrt{s} \approx 40~\mbox{GeV}$.

The measurement using approximately $30~\%$ of the data of 2002/2003 shows good agreement of the open beauty production cross section
$\sigma(b\bar{b}$ )with QCD calculations beyond NLO \cite{kido,boni} and the existing experimental results 
(Fig.~\ref{fig:final}) \cite{e789,e771,bb2000} within $1.5~\sigma$. The final measurement dealing with the total amount of statistics, can be used 
to reduce the theoretical uncertainties, originating mainly from uncertainties on the mass of the $b$--quark 
($m_{b} \in [4.5,5.0]~\mbox{GeV}$). 

Searches for exclusive $b$--decay search $B^{\pm} \rightarrow J/\psi K^{\pm}$, $B^{0} \rightarrow J/\psi K^{\pm}\pi^{\mp}$ and
$B^{0} \rightarrow J/\psi K^{0}_{S}$ are in progress, as well as searches for double semileptonic decays $b\bar{b} \rightarrow \mu\mu X$,
$b \rightarrow \mu\nu X$. These analyses will provide alternative $\sigma(b\bar{b})$ measurements and allow to cross check the
inclusive cross section measurement.

\end{document}